\documentclass[doublecol]{epl2} 

\begin{document}

\title{Patterns and bifurcations in low-Prandtl number Rayleigh-B\'{e}nard convection}

\author{Pankaj Kumar Mishra\inst{1} \and Pankaj Wahi\inst{2} \and Mahendra K. Verma\inst{1} }

\institute{                    
  \inst{1} Department of Physics, Indian Institute of Technology, Kanpur, India\\
  \inst{2} Department of Mechanical Engineering, Indian Institute of Technology, Kanpur, India
}


\date{\today}

\abstract{We present a detailed bifurcation structure and associated flow patterns for low-Prandtl number ($P=0.0002, 0.002, 0.005, 0.02$) Rayleigh-B\'{e}nard convection near its onset.   We use both direct numerical simulations and a 30-mode low-dimensional model for this study.  We observe that low-Prandtl number (low-P) convection exhibits similar patterns and chaos as zero-P convection~\cite{pal:2009}, namely squares, asymmetric squares, oscillating asymmetric squares, relaxation oscillations, and chaos.   At the onset of convection, low-P convective flows have stationary 2D rolls and associated stationary and oscillatory asymmetric squares in contrast to zero-P convection where chaos appears at the onset itself.  The range of Rayleigh number for which stationary 2D rolls exist decreases rapidly with decreasing Prandtl number.   Our results are in qualitative agreement with results reported earlier.}

\maketitle


Thermal convection is an important problem of classical physics, engineering, geophysics, atmospheric physics, and astrophysics.  As a consequence of its inherent nonlinearity, good understanding of convection eludes us even after sustained efforts for more than a century.  Convection in an arbitrary geometry is quite complex, so researchers have focussed on convection between two conducting parallel plates.  This idealized problem is called Rayleigh-B\'{e}nard convection (RBC). 

Convective flow in RBC is characterized by two non-dimensional parameters: the Rayleigh number $R$ which is the ratio of the destabilizing force to the stabilizing force, and the Prandtl number $P$ which is the ratio of the kinematic viscosity to the thermal diffusivity.  RBC shows a wide range of phenomena including instabilities, patterns, chaos, spatio-temporal chaos, and turbulence for various ranges of $R$ and $P$~\cite{chandrashekher}.  For low-Prandtl number (low-P) and zero-Prandtl number (zero-P) convection, the inertial term ${\bf u} \cdot \nabla {\bf u}$ becomes quite important and generates vertical vorticity.  As a result, the flow pattern becomes three-dimensional, and oscillatory waves along the horizontal axes are generated just near the onset of convection.  On the contrary, for large-Prandtl number (large-P) convection, vertical vorticity is absent near the onset and the two-dimensional (2D) rolls survive till large Rayleigh numbers~\cite{busse:1989}.  

   In this letter we investigate the bifurcation scenario for low-P convection and compare it with the recent  zero-P convection results of Pal {\em et al.}~\cite{pal:2009}. We will also explore the origin of chaos in low-P convection.   The relevant equations (under Boussinesq approximation) in nondimensionalized form are
\begin{eqnarray}
\partial_{t}(\nabla^{2}v_3) & = &\nabla^{4}v_{3}+R\nabla_{H}^{2}\theta \nonumber \\& & -\hat{e_3}\cdot\nabla\times[({\mathbf \omega} \cdot\nabla) {\mathbf v} -({\mathbf v} \cdot\nabla)\omega],\label{eq:u}\\ 
\partial_{t}\omega_{3}  & = & \nabla^{2}\omega_{3} + [({\mathbf \omega} \cdot\nabla)v_{3}-({\mathbf v} \cdot\nabla)\omega_{3}],\label{eq:omega}\\
P(\partial_t{\theta}+(\textbf{v}\cdot\nabla)\theta) & = & v_{3} + \nabla^{2}\theta,\label{eq:T}\\
\nabla\cdot {\bf v} = 0 \label{eq:incompressible}
\end{eqnarray}
where $\textbf{v}\equiv(v_1,v_2,v_3)$ is the velocity field, $\theta$ is the deviation in the temperature field from the steady conduction profile, $\bf{\omega} = \bf{\nabla}\times \textbf{v}$ is the vorticity field, $R=\alpha g (\Delta T) d^3/\nu \kappa$ is the Rayleigh number, $P=\nu/\kappa$ is the Prandtl number, $\hat{e}_3$ is the unit vector along the vertical (buoyancy) direction, and $\nabla_{H}^{2} = \partial_{xx}+\partial_{yy}$ is the horizontal Laplacian. Here, $\nu$ and $\kappa$ are fluid's kinematic viscosity and thermal diffusivity respectively, $d$ is the vertical distance between the two plates, and $\Delta T$ is the temperature difference between the horizontal plates.  For the above nondimensionalization, we used $d$ as the length scale, $d^2/\nu$ as the time scale, and $\nu \Delta T/\kappa$ as the temperature scale.  We consider {\it perfectly conducting and free-slip} conditions at the top and bottom plates and periodic boundary conditions along the horizontal directions. We will mostly use the reduced Rayleigh number $r=R/R_c$, where $R_c$ is the critical Rayleigh number, as the control parameter for our bifurcation analysis.
   
Instabilities and chaos in low-P and zero-P convection have been widely studied using analytical tools (e.g., approximation techniques), experiments, and numerical simulations. Busse~\cite{busse:1972} showed using perturbative analysis that for small Prandtl numbers, the 2D rolls become unstable as a result of oscillatory three-dimensional disturbances when the amplitude of the convective motion exceeds a critical value.  Busse and Bolton~\cite{busse:1984} argued that under free-slip boundary conditions, the stability of the 2D rolls exists only for Prandtl numbers above a critical value $P_c$, which is around 0.543.   Clever and Busse~\cite{clever:1974} extended the oscillatory instability analysis to no-slip boundary conditions and showed that the convective rolls are unstable for Prandtl numbers less than about 5. 

Siggia and Zippelius~\cite{siggia:1981} analyzed convective instabilities using amplitude equations and reasoned that for finite Prandtl numbers, the  vertical vorticity modes are important and the flow is three-dimensional. Fauve {\em et al.}~\cite{fauve:1987} investigated the origin of instabilities in low-P convection using the phase dynamical equations and argued that the instability always saturates into travelling waves as observed in experiments and numerical simulations. Jenkins and Proctor~\cite{Jenkins:1984} studied the transition from 2D rolls to square patterns in RBC for different Prandtl numbers using analytical tools.  Jones {\em et al.}~\cite{jones:1976} studied connection of the Prandtl number and the Nusselt number using perturbative techniques and numerical simulations.  Kumar {\em et al.}~\cite{kumar:1996} investigated  zero-P convection using a 6-mode model and showed the existence of wavy instability at the onset. They demonstrated that the saturation of the growing nonlinear roll solutions is due to the wavy instability.

Interesting experiments have been performed to explore the instabilities and chaos in convective flows near the onset. Willis and Deardorff~\cite{willis:1970} carried out experiments on air ($P=0.7$) and observed time dependence as a result of wavy instability. Rossby ~\cite{rossby:1969} demonstrated the existence of wavy instability in mercury ($P=0.02$). Krishnamurti~\cite{krishnamurti:1970} performed extensive convection experiments on a variety of fluids ($P=0.1-50$) and observed stationary and time-dependent three-dimensional patterns in her experiments; time dependent rolls appear at much lower Rayleigh numbers for low-P convection compared to that for large-P convection. 

Numerical simulations complement experiments in the investigation of instabilities and chaos in convection.   Lipps~\cite{lipps:1976} simulated convective flows in air ($P\approx 0.7$) and investigated various states including 2D rolls, time-periodic and aperiodic convection as a function of Rayleigh number and aspect ratio.  Bolton and Busse~\cite{bolton:1985} studied the stability of the steady convection rolls with respect to arbitrary  three-dimensional infinitesimal disturbances and observed stability of steady solutions for only a small regime of Rayleigh numbers and wavenumbers. Clever and Busse~\cite{clever:1990} simulated convective flows for $P=0.02$ with rigid boundary conditions using the Galerkin technique and observed asymmetric travelling wave patterns.  Meneguzzi {\em et al.}~\cite{meneguzzi:1987} performed three-dimensional convective simulations for $P=0.2$ fluid under stress-free boundary conditions, and for $P=0.025$ for no-slip boundary conditions.  For $P=0.2$ they observed stationary, periodic, biperiodic and chaotic regimes as the Rayleigh number is increased.  For the latter, they found stationary and time-periodic solutions only.  Thual~\cite{thual:1992} studied zero-P and low-P convection using numerical simulations for both no-slip and free-slip boundary conditions.  In his simulations he observed supercritical oscillatory instabilities, competition between two-dimensional rolls, square and hexagonal patterns, travelling and standing waves, and chaos for both zero-P and low-P convective flows. Ozoe and Hara~\cite{Ozoe:1995} performed  numerical simulations for Prandtl numbers  $P=0.001-0.1$ in a two-dimensional rectangular enclosure of aspect ratio 4 and observed that the time-dependent behaviour sets in at the onset of convection for $P=0.01$, thus decreasing the value of $P_c$ to much lower limit than that predicted by Busse and Bolton~\cite{busse:1984}.  
 
Recently Pal {\em et al.}~\cite{pal:2009} performed numerical simulations and bifurcation analysis using a low-dimensional model of zero-P convection under free-slip boundary conditions.  They could explain the origin of squares (SQ), asymmetric squares (ASQ), oscillating asymmetric squares (OASQ), relaxation oscillations with an intermediate square pattern (SQOR), and chaos using a series of bifurcations.  Pal {\em et al.}~\cite{pal:2009} start their analysis from square patterns at $ r =1.4 $.  Under reduction of $r$,  SQ bifurcates into asymmetric squares through a pitchfork bifurcation, which in turn transforms into oscillating asymmetric squares through a Hopf bifurcation.  OASQ  transforms to SQOR that subsequently becomes chaotic.  The chaotic attractor continues till the onset of convection.  There have been other attempts to study RBC using low-dimensional models  for other ranges of Prandtl numbers~\cite{low-d-highP}.

In this letter we perform numerical simulations and bifurcation analysis of low-P convection along similar lines as Pal {\em et al.}~\cite{pal:2009} to investigate the origin of patterns and chaos.  We also  carry out a comparative study of low-P convection with zero-P convection and show how the limiting behavior of zero-P convection is obtained as $P$ approaches zero.   

We perform direct numerical simulations (DNS) of convective flows for $P=0.02,  0.005, 0.002$ and $0.0002$ for $r$ ranging from 1 to 1.247  on $64^3$ rectangular grid.  We use a pseudo-spectral procedure with fourth-order Runge-Kutta as the time-stepping scheme. The aspect ratio of our simulations is $2\sqrt{2} : 2\sqrt{2} : 1$.  Eqs.~(\ref{eq:u}) and (\ref{eq:T}) provide us an estimate of $dt$ for the DNS ($dt \sim P \theta / v_3 \sim P/R$).  For low-P simulations this $dt$ is typically much smaller than that obtained using the CFL condition.  For example, for $P=0.0002$, $dt \sim 10^{-7}$ that makes numerical simulations very demanding.   In DNS we observe stationary 2D rolls, stationary asymmetric square patterns (ASQ), wavy rolls,  oscillatory asymmetric squares (OASQ), relaxation oscillation with squares (SQOR), and symmetric square patterns (SQ) for $P=0.02$.  A detailed description of the above phenomena in terms of bifurcation diagrams is described below.

  For a detailed bifurcation analysis of this problem, we construct a 30-mode low-dimensional model using the energetic modes of the DNS~\cite{pal:2009}. We pick 11 large-scale vertical velocity modes: $W_{101}$,$W_{011}$, $W_{112}$, $W_{211}$, $W_{121}$, $W_{301}$, $W_{103}$, $W_{013}$, $W_{031}$, $W_{202}$, $W_{022}$; 12 large-scale $\theta$ modes: $\theta_{101}$, $\theta_{011}$, $\theta_{112}$, $\theta_{211}$, $\theta_{121}$, $\theta_{301}$, $\theta_{103}$, $\theta_{013}$, $\theta_{031}$, $\theta_{202}$, $\theta_{022}$, $\theta_{002}$; and $7$ large-scale vertical vorticity modes: $Z_{110}$, $Z_{112}$, $Z_{211}$, $Z_{121}$, $Z_{220}$, $Z_{310}$, $Z_{130}$. The three subscripts here are the indices of the wavenumbers along the $x$, $y$, and $z$ directions.   All the Fourier modes, $30$ in total, are taken to be real. This model is inspired by the 13-mode model for zero-P convection~\cite{pal:2009}.  Next we investigate the origin of the convective flow patterns using a bifurcation diagram of this low-dimensional model following the approach of Pal {\em et al.}~\cite{pal:2009}. The fixed point(s) are generated numerically using the Newton-Raphson  method for a given $r$, and they are subsequently continued using a fixed arc-length based continuation scheme~\cite{chatterjee:2005} for the neighbouring $r$ values. The stability of the fixed points and the parameter values corresponding to bifurcations are ascertained through an eigenvalue analysis of the Jacobian.  New branches of steady solutions (fixed points at pitchfork, while periodic solutions at Hopf) are generated by calculating and continuing these solutions. The stability of these new solutions is ascertained analogously to obtain subsequent bifurcations and new branches of solutions.  A MATLAB package named MATCONT was used for some of the above analysis.

\begin{figure}[ht]
  \begin{center}
  \includegraphics[width=1.0\columnwidth]{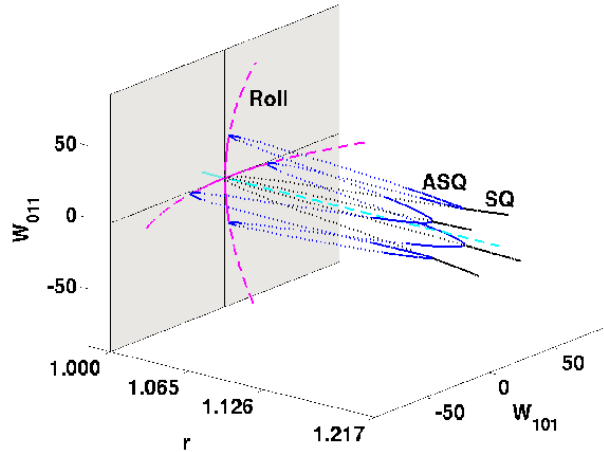}
  \end{center}
  \caption{For $P=0.02$, the three dimensional view of the bifurcation diagram computed from the low-dimensional model exhibiting stable fixed points (solid curves), and unstable fixed points (dotted and dashed curves). The purple solid and dashed lines represent the stable and unstable 2D roll solutions respectively.  The black, blue, and cyan curves depict stationary squares (SQ), asymmetric stationary squares (ASQ), and conduction state respectively.}
  \label{fig:bifurc3D_P0p02}
  \end{figure}

The state space or the phase space of our low-dimensional model is 30 dimensional. However among all the modes, $W_{101}, W_{011}, \theta_{101}$, and  $\theta_{011}$ are the most important ones;  the behaviour of the rolls along the $x$ and the $y$ directions depends critically on these modes. In Fig.~\ref{fig:bifurc3D_P0p02} we plot the numerical values of the fixed points $W_{101} $ and $W_{011}$ for $P=0.02$ as a function of $r$. The stable fixed points are shown as solid lines, while the unstable ones are shown as dotted and dashed lines. At $r=1$, the conduction state (cyan curve) becomes unstable and bifurcates into four stable 2D rolls (shown as purple curves) and four unstable symmetric squares (SQ, shown as black dotted lines) through a codimension-2 pitchfork bifurcation associated with double zero eigenvalues. The stable 2D rolls along the $x$ axis have $W_{011} = 0$, while the rolls along the $y$ axis have $W_{101} = 0$.  The unstable  symmetric square (SQ) solutions satisfy $|W_{101}| = |W_{011}|$.   With an increase of $r$, the stable 2D rolls bifurcate into stable patterns called asymmetric squares (ASQ)  shown as the solid blue lines in Fig.~\ref{fig:bifurc3D_P0p02}.  These ASQ patterns lose stability at some point through a Hopf bifurcation. With a further increase in $r$, the unstable branch of ASQ solutions (dotted blue curves) regains stability through an inverse Hopf bifurcation resulting in stable ASQ solutions (solid blue curves). These stable ASQ solutions subsequently meet the unstable SQ patterns (represented by dotted black curves originating from $r=1$) in yet another inverse bifurcation, and stable SQ patterns (solid black curves) are formed.  These stable SQ patterns subsequently lose stability beyond $r=1.247$ and complex chaotic attractors are generated, however an analysis in this regime is beyond the scope of this letter.  This sequence of fixed points remains the same for all finite but small values of $P$. However, the range of $r$ corresponding to these solutions changes with $P$.

\begin{figure}[ht]
  \begin{center}
  \includegraphics[width=1.0\columnwidth]{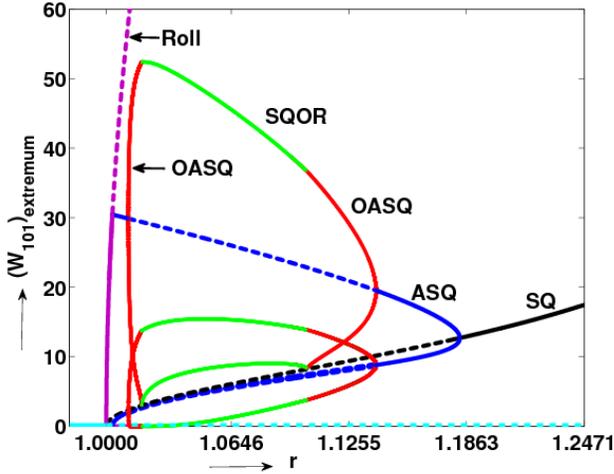}
  \end{center}
  \caption{Bifurcation diagram of the low-dimensional model for $P=0.02$ in the range  $0.95 \leq r \leq 1.247$. The stable branches corresponding to conduction state, 2D rolls, stationary squares (SQ) and stationary asymmetric squares (ASQ) are represented by solid cyan, solid purple, solid black, and solid blue curves respectively. The red and green curves depict the extrema of oscillatory asymmetric squares (OASQ) and relaxation oscillation with intermediate square patterns (SQOR) respectively.  The dashed curves represent unstable solutions.}
  \label{fig:bifurc2D_P0p02}
  \end{figure}
    
We will describe the above mentioned fixed points and associated time-dependent patterns using a bifurcation diagram in the range of $0.95 \le r \le1.247$. Figure~\ref{fig:bifurc2D_P0p02} illustrates the bifurcation diagram for $P=0.02$ where we plot the positive value of $(W_{101})_{extremum}$ as a function of $r$.  At $r=1$ the conduction state bifurcates to stationary 2D rolls (purple curve) through a codimension-2 pitchfork bifurcation.  As described above unstable stationary SQ (shown as dashed black curve) with $W_{101} = \pm W_{011}$ are also generated as a result of this bifurcation. As $r$ is increased further, at  $r \approx 1.0035$, the stationary 2D rolls bifurcate to ASQ patterns through another pitchfork bifurcation (solid blue curves).   Subsequently, at $r \approx 1.0114$, ASQ patterns bifurcate to limit cycles (red curves) through a Hopf bifurcation.    These limit cycles represent oscillatory asymmetric square patterns (OASQ).   Note that the limit cycles are quite close to the axes.   The oscillatory patterns corresponding to these limit cycles form  standing waves along the roll axis that have been  discussed earlier by Thual~\cite{thual:1992}.   The time-period and amplitude of these limit cycles increase as $r$ is increased.  At $r \approx 1.0184$, these limit cycles form homoclinic orbits of the unstable SQ fixed points (saddles) originating from $r=1$.  

 Beyond $r \approx 1.0184$, there is a smooth transition from homoclinic orbits to regular limit cycles that correspond to  relaxation oscillations between two roll solutions along the $x$ and the $y$ directions with intermediate square patterns (SQOR, illustrated as green curves in the bifurcation diagram). The SQOR patterns and all subsequent patterns generated at higher $r$ values are similar to that observed by Pal {\em et al.}~\cite{pal:2009} for zero-P convection.  In brief, the limit cycles corresponding to SQOR form another set of homoclinic orbits at $r\approx 1.1034$ after which each of the homoclinic orbits bifurcate into two separate limit cycles. These limit cycles (OASQ  shown as red curves) diminish in size as $r$ is increased until they transform to stable fixed points (ASQ shown as blue curves) through an inverse Hopf bifurcation.   As $r$ is increased further, at $r \approx 1.183$, the stable ASQ branches meet the unstable SQ branch (the dashed black curve) resulting in stabilization of the SQ pattern
(solid black curves) as a consequence of an inverse pitchfork bifurcation.  Thus, convective flow for $P=0.02$ exhibits SQ, ASQ, OASQ, and SQOR patterns that are common to zero-P convection~\cite{pal:2009}.   In addition, stationary 2D rolls and associated stationary and oscillating ASQ are observed near the onset of convection.   The chaotic attractors {\em Ch1, Ch2}, and  {\em Ch3} of zero-P convection~\cite{pal:2009} are absent for $P=0.02$.   In the following discussion we will consider the bifurcation diagrams for lower Prandtl numbers ($P=0.005, 0.002, 0.0002$) for which chaotic attractors are observed.
 
We performed DNS for $P=0.02$ and observed similar behaviour as the low-dimensional model.  A close similarity between the model and the DNS is due to the fact that the modes for the low-dimensional models were chosen from DNS runs by identifying the energetic modes.   This exercise is possible near the onset of convection where only a limited number of modes are excited. Comprehensive analysis of DNS and low-dimensional model will be presented elsewhere.

\begin{figure}[ht]
  \begin{center}
  \includegraphics[width=1.0\columnwidth]{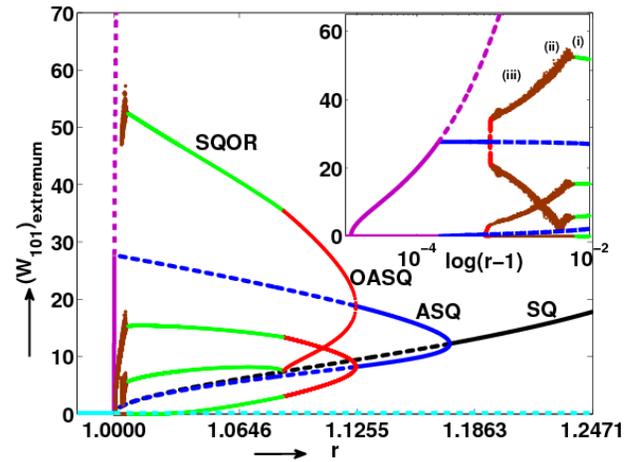}
  \end{center}
  \caption{Bifurcation diagram of the low-dimensional model for $P=0.005$ in the range $0.95 \leq r \leq 1.247$.   The color convention is same as that for $P=0.02$ (Fig.~\ref{fig:bifurc2D_P0p02}).  The chaotic attractors are shown in brown colour.  A zoomed view of the bifurcation diagram for the chaotic regime is shown in the inset where the $x$ axis is chosen as $\log(r-1)$ to highlight the behaviour near $r=1$. The three attractors {\em Ch1, Ch2} and {\em Ch3} are shown as (i), (ii), and (iii) respectively. }
  \label{fig:bifurc2D_P0p005}
  \end{figure}
  
\begin{figure}[ht]
  \begin{center}
  \includegraphics[width=1.0\columnwidth]{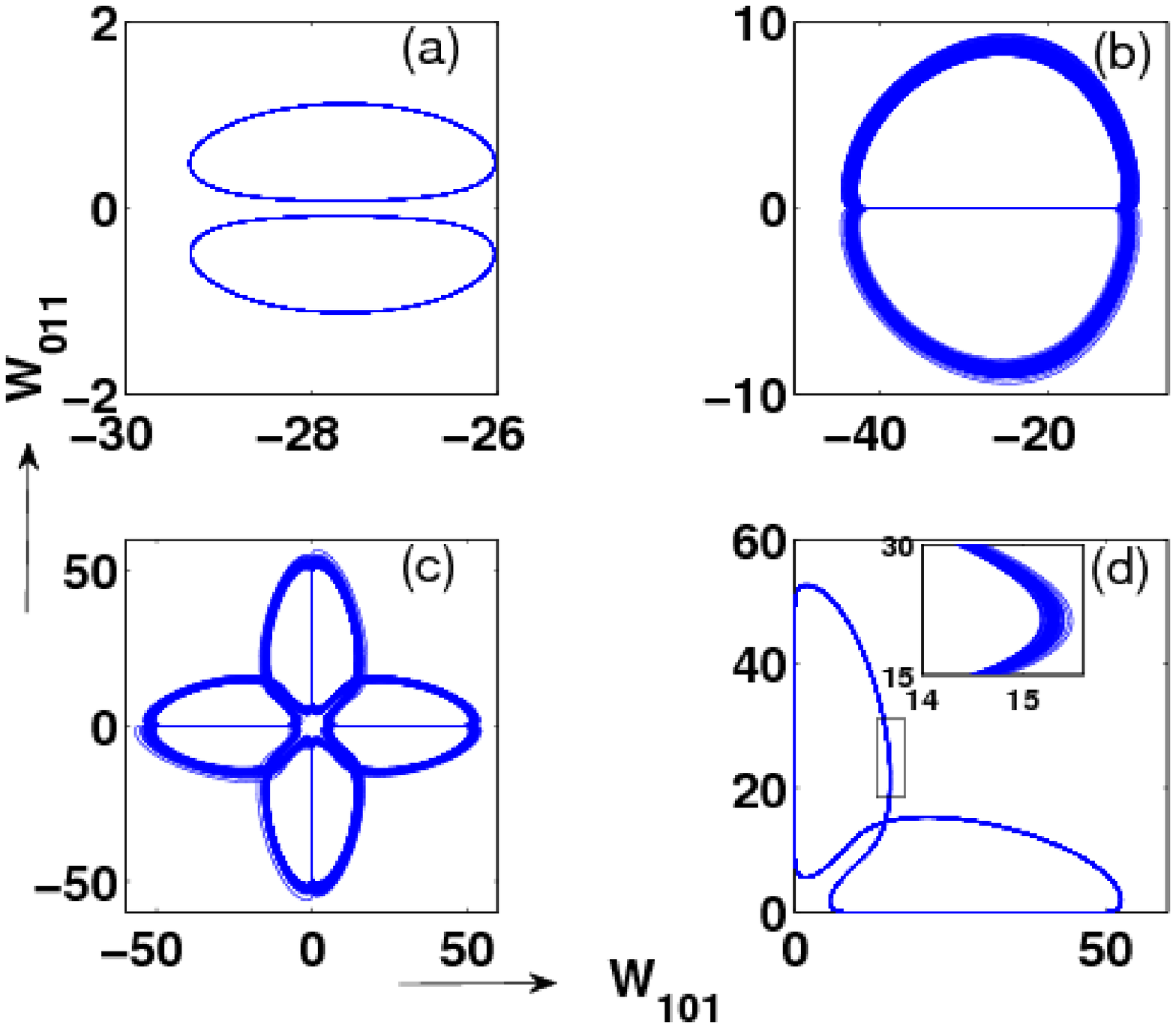}
  \end{center}
  \caption{Phase space projections of the attractors on the $W_{101}$-$W_{011}$ plane obtained from the low-dimensional model for $P=0.005$ near $r=1$: (a) two of the limit cycles at $r=1.0006844$; (b) the chaotic attractor {\em Ch3} at $r=1.0023$;  (c) the chaotic attractor {\em Ch2}  at $r=1.0053$;  (d) the chaotic attractor {\em Ch1} at $r=1.0064$, with the inset depicting the chaotic nature of the attractor.} 
 \label{fig:model_chaotic_P0p005}
  \end{figure}

After the discussion on the bifurcation scenario for $P=0.02$, we turn to lower Prandtl numbers $P=0.005, 0.002$,  and $0.0002$.  The bifurcation diagram for $P=0.005$, shown in Fig.~\ref{fig:bifurc2D_P0p005}, is  similar to the corresponding figure for $P=0.02$ with a major difference that for $P=0.005$,  the chaotic attractors {\em Ch1, Ch2}, and {\em Ch3} of zero-P convection~\cite{pal:2009} appear near the onset in the band of $r = 1.000685-1.0068$.  As we increase $r$ from 1, we observe 2D straight rolls, ASQ, and OASQ just like $P=0.02$, however, at lower $r$ values (see Table~\ref{Table1}).    The phase space projection on the $W_{101}$-$W_{011}$  plane of two of the limit cycles corresponding to OASQ are shown in figure~\ref{fig:model_chaotic_P0p005}(a).  Subsequently these limit cycles appear to approach their basin boundary (the horizontal axis of the figure), and  the system becomes chaotic. This phenomenon is due to a ``gluing bifurcation"~\cite{gluing}, and it could be related to the  ``attractor merging crisis''~\cite{Ott:book}.  The resulting chaotic attractor, whose phase space projection is shown in Fig.~\ref{fig:model_chaotic_P0p005}(b),  is same as the {\em Ch3} attractor of zero-P convection~\cite{pal:2009}.  As $r$ is increased further, the various {\em Ch3} attractors merge together in yet another crisis to form a single large chaotic attractor {\em Ch2}.  At a larger $r$ value, the {\em Ch2} attractor breaks into four separate chaotic attractors {\em Ch1}. Phase space projections of the chaotic attractors {\em Ch2} and {\em Ch1} for $P=0.005$ are shown in Figs.~\ref{fig:model_chaotic_P0p005}(c) and ~\ref{fig:model_chaotic_P0p005}(d) respectively. The {\em Ch1} chaotic attractors become regular for $r \ge 1.0068$ giving rise to the SQOR limit cycles. The subsequent patterns and the bifurcation diagram are same as those for $P=0.02$. Note that the {\em Ch1} chaotic attractors are generated as a result of ``homoclinic chaos''~\cite{pal:2009}.   
\begin{table}[ht]
\caption{Table depicting the reduced Rayleigh number $r_p$ at which the 2D rolls bifurcate to ASQ, $r_t$ where ASQ patterns bifurcate to limit cycles, and the frequency $\omega$ of the limit cycle at the Hopf bifurcation point.  These values are computed for Prandtl numbers $P=0.02, 0.005, 0.002$ and 0.0002.} 

\begin{tabular}{ c c c c} \hline
$P$ & $ r_p $&  $r_t$ & $\omega$  \\ \hline
0.02& 1.0035& 1.01139 & $1.15\times10^{-3}$\\
0.005 &1.000179&1.000683& $2.9\times10^{-5}$\\
0.002 & 1.000042& 1.00012 & $1.15\times10^{-5}$ \\
0.0002 & 1+$2.5\times 10^{-7}$ & 1.000018 & $1.15\times10^{-7}$\\ \hline
\end{tabular}
\label{Table1}
\end{table}

\begin{figure}[ht]
  \begin{center}
 \includegraphics[width=1.0\columnwidth]{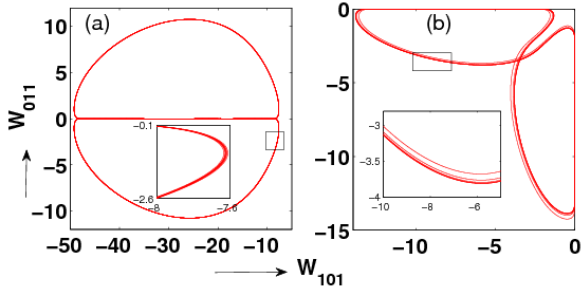}
  \end{center}
  \caption{(a) Phase space projection of the chaotic attractor {\em Ch3}  on the $W_{101}$-$W_{011}$ plane obtained from DNS with $P=0.005$ and $r=1.0023$;  (b) Phase space projection of the chaotic attractor {\em Ch1}  on the $W_{101}$-$W_{011}$ plane obtained from DNS with $P=0.005$ and $r=1.0068$.  The insets show an enlarged view of the boxed region.}
  \label{fig:dns_chaotic_P0p005}
  \end{figure}

The bifurcation scenario presented above for $P=0.005$ has  all the features (including chaos) of zero-P convection along with the stable 2D rolls and associated ASQ and OASQ near the onset.  Note that for $P=0.02$, the window of chaos ({\em Ch1} to {\em Ch3}) is absent as the homoclinic intersection point of the left OASQ limit cycle precedes  the gluing bifurcation point.   The patterns and chaotic attractors observed in the low-dimensional model for $P=0.005$ are also observed in DNS.   Phase space projections of the chaotic attractors ({\em Ch3} and {\em Ch1}) obtained in DNS are shown in Figs.~\ref{fig:dns_chaotic_P0p005}(a) (for $r=1.0023$) and~\ref{fig:dns_chaotic_P0p005}(b) (for $r=1.0068$) respectively. The chaotic attractors obtained from the DNS data are not thick, however an enlarged view of the attractor in the inset shows that the attractor is chaotic.    

\begin{figure}[ht]
 \begin{center}
  \includegraphics[width=1.0\columnwidth]{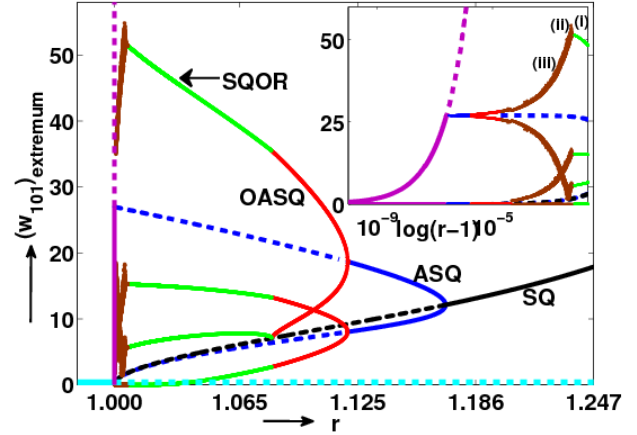}
\end{center}
  \caption{Bifurcation digram of the low-dimensional model for $P=0.0002$ in the range $0.95 \leq r \leq 1.247$.   The color convention is same as that for $P=0.005$ (Fig.~\ref{fig:bifurc2D_P0p005}).  A zoomed view of the bifurcation diagram for the chaotic regime is shown in the inset where the $x$ axis is chosen as $\log(r-1)$ to highlight the behaviour near $r=1$.}
  \label{fig:bifurc2D_P0p0002}
 \end{figure}

When we lower the value of the Prandtl number even further from $P=0.005$, the stationary 2D rolls and the associated ASQ patterns occur even nearer to $r=1$ as evident from the entries of Table~\ref{Table1}.  It can be noted from Fig.\ref{fig:bifurc2D_P0p0002} that the windows of 2D rolls and ASQ for $P=0.0002$ is very small.  The inset of Fig.~\ref{fig:bifurc2D_P0p0002} shows that the 2D rolls and ASQ occur for $1 < r < 1+2.5 \times 10^{-7}$ and $ 1+2.5 \times 10^{-7} < r < 1+1.8 \times 10^{-7}$ respectively.    Consequently it would be very difficult to observe these 2D rolls and the related ASQ patterns in experiments and in DNS; a small experimental or numerical noise will be sufficient to push the system out of this narrow region to the chaotic region {\em Ch1-Ch3}, which is a feature of zero-P convection.  Thus our study indicates that zero-P convection is an appropriate limiting case of low-P convection in terms of bifurcations near the onset.  For practical purposes, the bifurcation diagram for very low-P convection is topologically equivalent to zero-P convection.

It can be seen from Table~\ref{Table1} that the branch points corresponding to ASQ ($r_p$) and OASQ  ($r_t$)  asymptotically approach $r=1$ as $P \rightarrow 0$.  In Table~\ref{Table1} we also list the imaginary part of the eigenvalue ($\omega$)  of the stability matrix at the Hopf bifurcation ($r = r_t$).   We  observe that $r_p-1$, $r_t-1$, and $\omega$ appear to vary approximately as $P^2$ with prefactors around 10, 30 and 3 respectively.  The $P^2$ dependence is consistent with the theoretical predictions of  Busse~\cite{busse:1972} that $r_t -1 \ge 0.310 P^2$ for free-slip boundary conditions.  However the multiplying factor of Busse and our model differ by an order of magnitude.  Also, Busse and Bolton~\cite{busse:1984}  predict an absence of stable 2D rolls for $P< P_c = 0.543$.  Our analysis however indicates stable 2D rolls for nonzero $P$, with the range of $r_p-1$ decreasing rapidly with the lowering of $P$.  Earlier Krishnamurti~\cite{krishnamurti:1970} had found stable 2D rolls  for mercury ($P=0.025$) near the onset ($R  \approx (2.3 \pm 0.1) \times 10^3$) in her experiments with no-slip boundary conditions, thus indicating that $P_c$ predicted by Busse and Bolton~\cite{busse:1984} is an overestimate. Regarding $\omega$, Busse~\cite{busse:1972} predicts that $\omega \sim P$ in contrast to our finding that $\omega \sim P^2$ near the onset of convection. In a related work on convection, Mercader {\em et al.}~\cite{mercader:2005} have shown that the relationship between $\omega$ and $P$ depends critically on the aspect ratio and they have reported both linear and quadratic dependence. These issues require a more careful and detailed theoretical investigation.

 A comparison of the above results with the bifurcation diagram of zero-P convection of Pal {\em et al.}~\cite{pal:2009} reveal strong similarities.  The bifurcation scenario from SQ to SQOR for the larger $r$ values are topologically equivalent.  We observe two major differences between the bifurcation diagrams of zero-P convection and low-P convection:  (a) no chaos for $P \ge 0.02$ near the onset;  (b) existence of stable 2D rolls and associated ASQ and OASQ patterns near the onset.  For $P=0.02$  we find only stationary and time-periodic solutions near the onset which is in agreement with the results of Meneguzzi {\em et al.}~\cite{meneguzzi:1987} for $P=0.025$ and Thual~\cite{thual:1992} for $P=0.02$ and 0.2. However, for $P \le 0.005$ we observe chaotic attractors {\em Ch1, Ch2} and {\em Ch3} between OASQ (closer to $r=1$) and SQOR.   The stationary 2D rolls and associated ASQ and OASQ patterns are observed in a narrow window between the conduction state and the chaotic states.

 In conclusion, we perform a bifurcation analysis for low-P convection near the onset using DNS and a related low-dimensional model.  The results of the low-dimensional model are in good agreement with those of DNS since the model is derived using DNS by choosing most of the relevant modes.  We have chosen aspect ratio of  $\Gamma_x = \Gamma_y= 2 \sqrt{2}$.   We observe that the bifurcation diagram for low-P convection is very similar to the zero-P convection reported by Pal {\em et al.}~\cite{pal:2009}, except near the onset of convection where 2D stationary rolls, and stationary and oscillatory asymmetric squares are observed for nonzero Prandtl numbers. The range of Rayleigh numbers for which 2D rolls and associated ASQ and OASQ are observed shrinks rapidly ($\sim P^2$) as $P$ is decreased.  This result is in qualitative agreement with the results of Busse~\cite{busse:1972}.
For $P \le 0.0002$, the range of reduced Rayleigh numbers for which the stationary 2D rolls could be observed is too narrow ($< 10^{-7}$) to be observed in experiments or in DNS.   Our results predict the critical Prandtl number $P_c$ to be much lower than that predicted by Busse and Bolton~\cite{busse:1984}.

 In the present letter we also report the origin of chaos near the onset of convection for low-P convective flows.  Here chaos appears through a series of crisis.  The same set of chaotic attractors also appear in zero-P convection~\cite{pal:2009}.  Thus the bifurcation diagrams for low-P convection ($P \le 0.005$) have significant similarities with zero-P convection, and zero-P convection is an asymptotic limit of low-P convection.   Our bifurcation analysis provides very useful insights into the origin of patterns and chaos for low-P convection.  An extension of the present work to different aspect ratios, and to low-dimensional models with a larger set of modes is in progress.

We thank Krishna Kumar, Stephan Fauve, Michael Proctor,  Supriyo Paul and Pinaki Pal for very useful discussions and comments.  We also thank creators of MATCONT, using which part of the bifurcation diagrams were created, and to Computational Research Laboratory, India for providing access to the supercomputer EKA where some of the DNS runs were performed.  This work was supported by a research grant of the Department of Science and Technology, India as Swarnajayanti fellowship to MKV.


\end{document}